\documentclass[11pt,twoside]{article}

\usepackage{asp2006}
\usepackage{epsf}
\usepackage{lscape}
\usepackage{graphicx}

\begin{document}
\title{Light Curve Analysis of the Hybrid SdB Pulsators\\HS\,0702+6043 and 
HS\,2201+2610}
\author{R. Lutz,$^1$ S. Schuh,$^1$ R. Silvotti,$^2$ S. Dreizler,$^1$
E. M. Green,$^3$\\G. Fontaine,$^4$ T. Stahn,$^5$ S. D. H\"ugelmeyer,$^1$ and 
T.-O. Husser$^1$ }
\affil{$^1$Institut f\"ur Astrophysik, Georg-August-Universit\"at G\"ottingen,
  Friedrich-Hund-Platz 1, 37077 G\"ottingen, Germany\\
$^2$INAF-Osservatorio Astronomico di Capodimonte, via Moiariello 16, 
80131 Napoli, Italy\\
$^3$Steward Observatory, University of Arizona, 933 North Cherry Ave.,
Tucson, AZ 85721, USA\\
$^4$D{\'e}partement de Physique, Universit{\'e} de 
Montr{\'e}al, C.P. 6128, Succursale Centre-Ville, Montr{\'e}al, 
Qu{\'e}bec, H3C\,3J7, Canada
$^5$Max-Planck-Institut f\"ur Sonnensystemforschung, Max-Planck-Stra\ss e 2,
  37191 Katlenburg-Lindau, Germany}

\begin{abstract} 
We present the detection of low-amplitude, long-period $g$-modes in two 
individual sdBV stars which are known to be $p$-mode pulsators. Only few of 
these hybrid objects, showing both $p$- and $g$-modes, are known today. We 
resolve the $g$-mode domain in HS\,0702+6043 and add HS\,2201+2610 to the list 
of hybrid pulsators. To discover the low-amplitude $g$-modes, a filtering 
algorithm based on wavelet transformations was applied to denoise 
observational data.
\end{abstract}

\section{Introduction}   
Subdwarf B stars are evolved and compact objects which are thought to be core 
helium burning. They populate the extreme horizontal branch (EHB) at effective 
temperatures of $20\,000$\,--\,$40\,000$~K and surface gravities 
$\log (g/\rm{cm\,s^{-2}})$ between 5.0 and 6.2. Variable sdB stars (sdBVs) 
can be divided into the following subclasses: The $p$-mode pulsators show 
higher amplitudes and shorter periods at higher temperatures. 
\citet{rl_kil97} discovered EC 14026-2647 as the class prototype. The $g$-mode 
pulsators have lower amplitudes and longer periods at lower temperatures and 
the class prototype PG\,1716+426 was discovered by \citet{rl_gre03}. Hybrid 
pulsators show both $p$- and $g$-modes. These are particularly exciting 
objects, since the two mode types probe different regions in the star. Both 
mode types are thought to be driven by the same $\kappa$-mechanism 
\citep{rl_cha97,rl_fon03,rl_jes06}. Figure 1 shows the location of some sdB 
pulsators in the $\log g$\,--\,$T_{\rm eff}$ diagram.
\begin{figure}[!ht]
\centering
\includegraphics[scale=0.45,angle=-270]{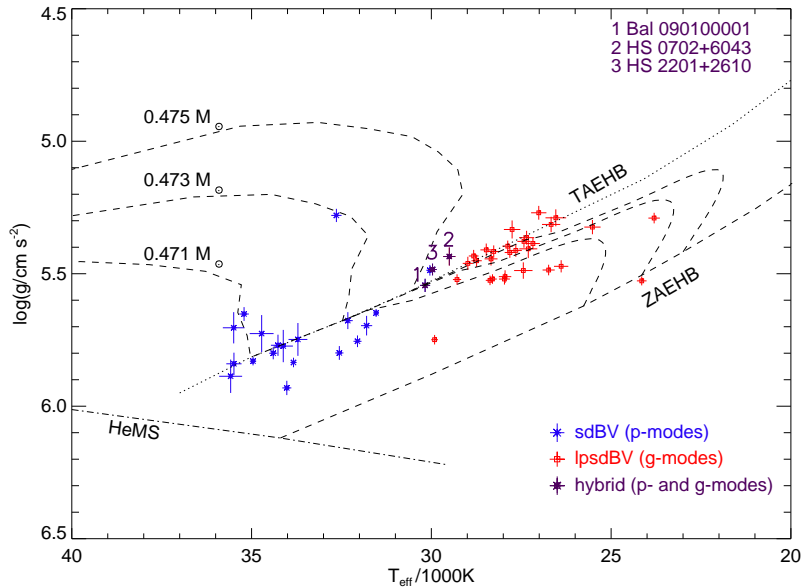}
\caption{Location of some known sdB pulsators in the $\log g$-$T_{\rm eff}$
  diagram. Evolutionary tracks \citep{rl_dor93} for three stellar masses are 
  indicated, as well as helium main sequence (HeMS) and zero age (ZAEHB) and 
  terminal age (TAEHB) extreme horizontal branches. The positions (taken from 
  E. M. Green and G. Fontaine; for a discussion see also Green et al., these 
  proceedings) clearly reveal two instability regions. The three known hybrid 
  pulsators are labelled 1, 2 and 3.}
\end{figure}
\section{Observations, Data Reduction and Wavelet Filter}
The HS\,0702+6043 photometric data have been taken at the Calar Alto 
Observatory in January 2005 with the 2.2\,m telescope equipped with CAFOS, 
using a B filter. From the whole ten day data set we used the best six nights 
($\approx$ 56 hours) for a further analysis after subtraction of the 
$p$-modes. The HS\,2201+2610 data have also
been taken at Calar Alto Observatory in September and November 2006 with
the same setup as above. Five nights of observation yield a total of about 26
hours of high S/N photometric data. We reduced the data in a standard way,
using the aperture photometry package TRIPP \citep{rl_sch03}.
To denoise our HS\,0702+6043 data, we applied the so-called $\grave a$ $trous$ 
algorithm \citep{rl_fli97} which is based on wavelet transformations. The
basic procedure of the filter is to decompose the noisy light curve into 
different scales, i.e.\ different frequency bands, and to compare 
transformation coefficients to a simulated pure noise light curve. Setting a 
threshold value during the reconstruction therefore produces a filtered and 
denoised light curve.
\section{Frequency Analysis of HS\,0702+6043 and HS\,2201+2610}
The following frequency spectra (Lomb-Scargle-Periodograms, LSPs) have been 
generated using the Scargle algorithm, because it can handle unevenly sampled 
data. The confidence levels (derived from false alarm probabilities) are 
based on ten thousand white noise simulations. Figure 2 shows the window
functions. The hybrid character of HS\,0702+6043 is described in 
\citet{rl_sch06} and with the 2005 data presented here, we are able 
to provide a better resolution of the $g$-mode domain. The results of our 
frequency analysis are summarized in the top part of Table 1. Figure 3 
displays the $g$-mode regime. In our 2006 data of HS\,2201+2610 we detect a 
long-period variation shown in Figure 4. Other data provided by one of us 
(R. S.) also show this variation. Our frequency analysis is summarized in the 
bottom part of Table 1. In addition to the hybrid behaviour, the interest for 
HS\,2201+2610 increased a lot recently due to the presence of an orbiting 
planet (\citealt{rl_sil07}, and these proceedings).
\section{Results}
We resolve three gravity modes in the hybrid pulsator HS\,0702+6043 and 
detect a long-period variation in HS\,2201+2610. We therefore propose to add
the latter to the list of hybrid pulsators.  
\begin{table}[!ht]
\caption{Frequencies appearing in HS\,0702+6043 (top) and HS\,2201+2610 
  (bottom). \textit{Top}: We estimate the errors in the frequency to 
  be 0.9\,$\mu$Hz (main FWHM in the window function) and having an 
  accuracy of half a mmi in the amplitudes at a detection limit of 
  0.8\,mmi. \textit{Bottom}: We estimate the errors in the frequency 
  to be 0.5\,$\mu$Hz and having a detection limit of 1.1\,mmi. We used (f4) 
  and (f5) in the prewhitening, because low-amplitude signals in this 
  frequency range were also found in other runs \citep{rl_sil02}.}
\smallskip
\begin{center}
{\small
\begin{tabular}{ccrrrr}
\tableline
\noalign{\smallskip}
Object&ID & Frequency & Period & Amplitude & type\\
&   & [$\mu$Hz] & [s] & [mmi] & \\
\noalign{\smallskip}
\tableline
\noalign{\smallskip}
HS\,0702+6043&f1 & 2753.9 & 363.1 & 21.7 & $p$\\
&f2 & 2606.1 & 383.7 & 4.6 & $p$\\
&f3 & 271.7 & 3680.8 & 1.8 & $g$\\
&f4 & 318.1 & 3144.2 & 1.3 & $g$\\
&f5 & 206.3 & 4847.0 & 0.9 & $g$\\
\tableline
\noalign{\smallskip}
HS\,2201+2610&f1 & 2861.2 & 349.5 & 6.6 & $p$\\
&f2 & 2823.7 & 354.2 & 3.3 & $p$\\
&f3 & 2880.8 & 347.1 & 1.5 & $p$\\
&(f4) & 2725.9 & 366.9 & 0.8 & $p$\\
&(f5) & 2906.3 & 344.1 & 0.4 & $p$\\
&f6 & 307.2 & 3255.6 & 1.5 & $g$\\
\noalign{\smallskip}
\tableline
\end{tabular}
}
\end{center}
\end{table}
\begin{figure}[!ht]
\centering
\includegraphics[scale=0.20,angle=-270]{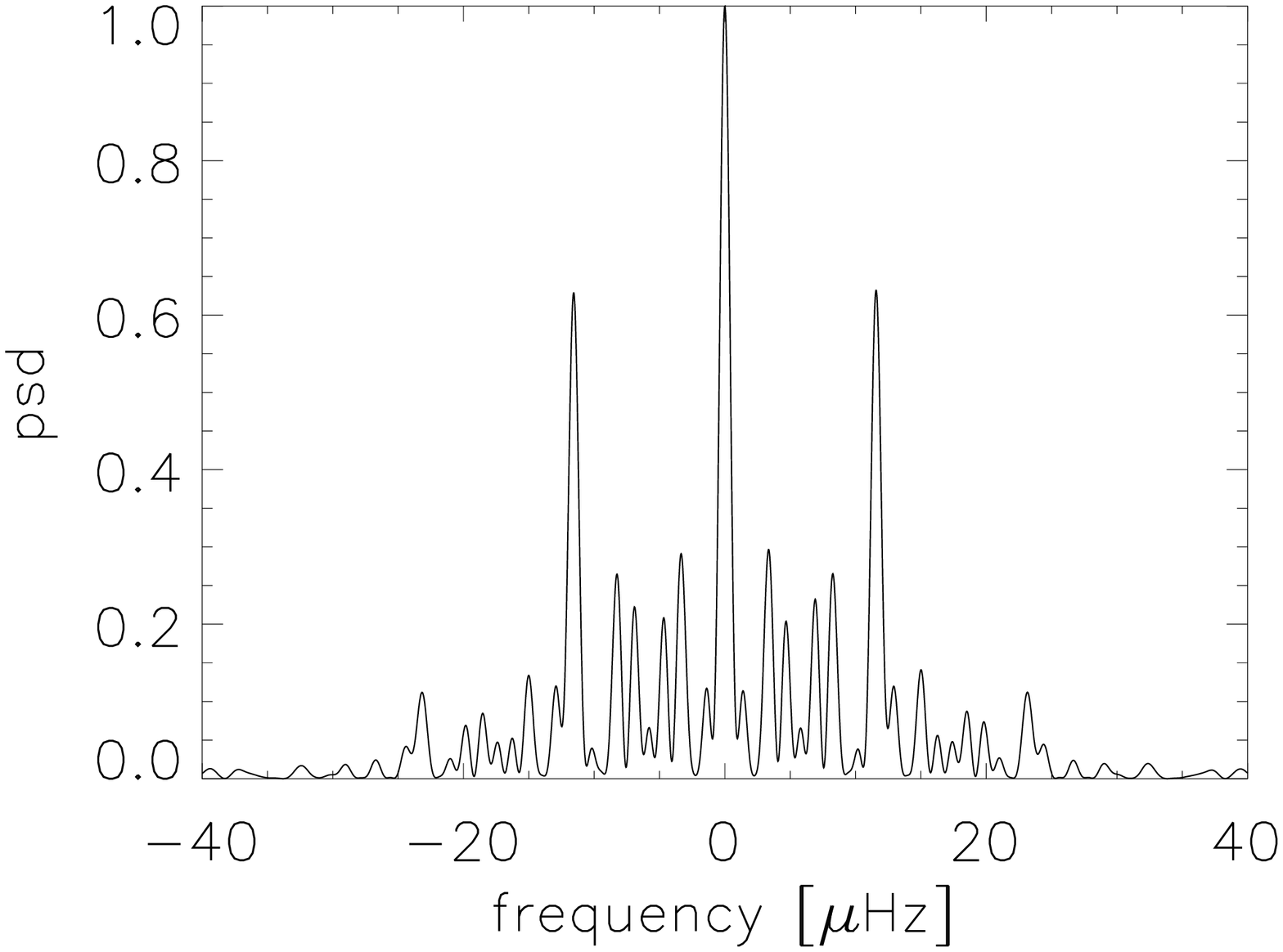}
\includegraphics[scale=0.20,angle=-270]{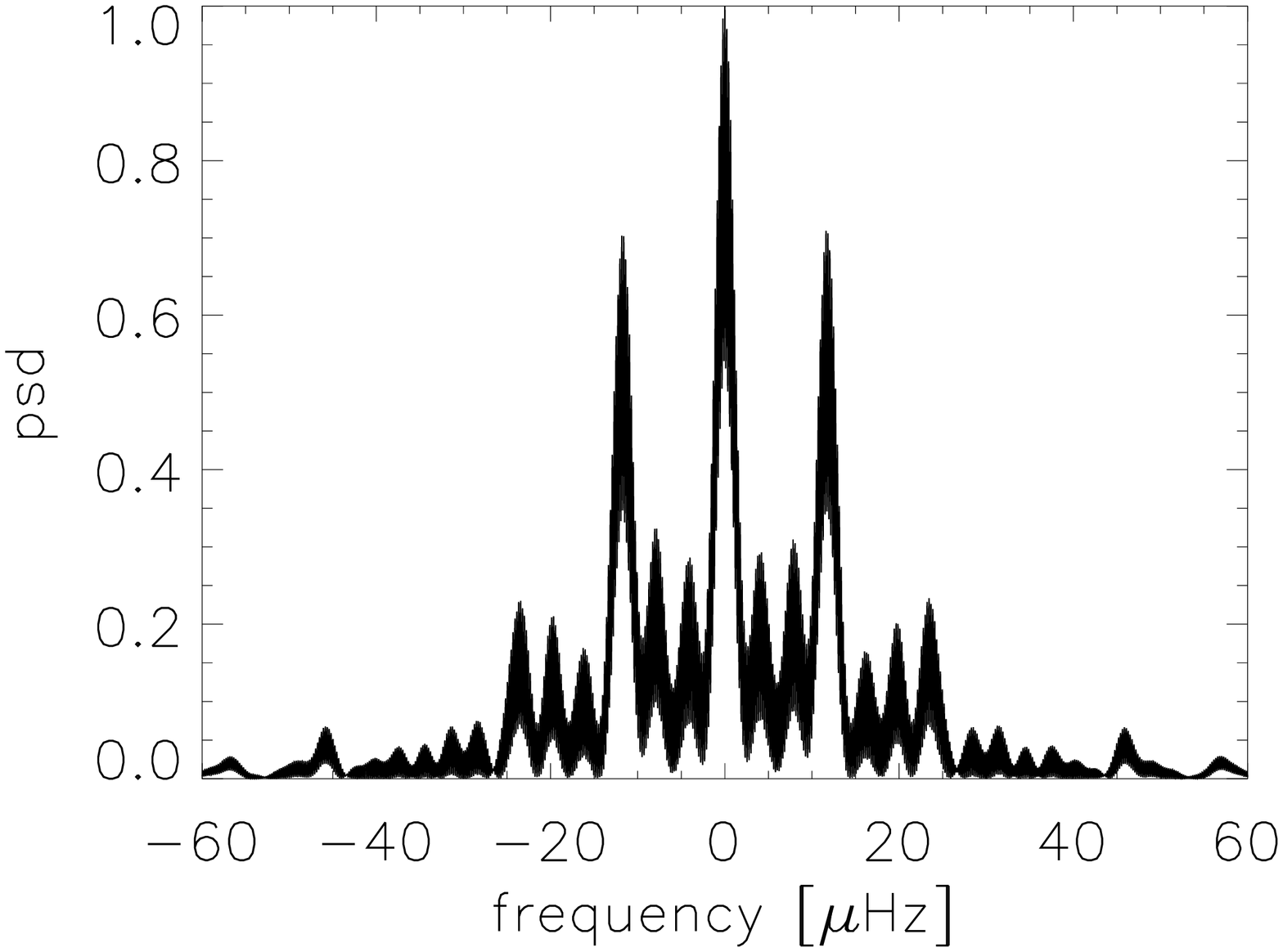}
\caption{Window functions of our HS\,0702+6043 (left) and HS\,2201+2610
  data (right).}
\end{figure}
\begin{figure}[!ht]
\centering
\includegraphics[scale=0.27,angle=-270]{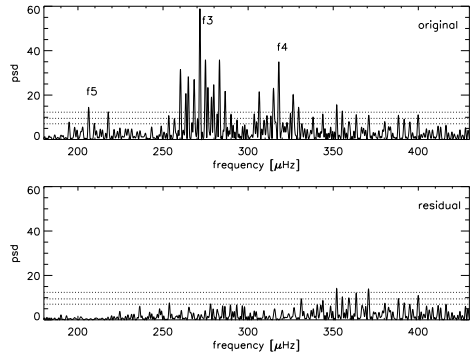}
\caption{LSPs of the $g$-mode regime of HS\,0702+6043 after subtraction 
of two $p$-modes (top). Simultaneously subtracting the $g$-mode 
frequencies f3, f4 and f5 results in the residual LSP (bottom). 
Horizontal lines indicate confidence levels of 3\,$\sigma$, 
2\,$\sigma$ and 1.5\,$\sigma$.}
\end{figure}
\begin{figure}[!ht]
\centering
\includegraphics[scale=0.27,angle=-270]{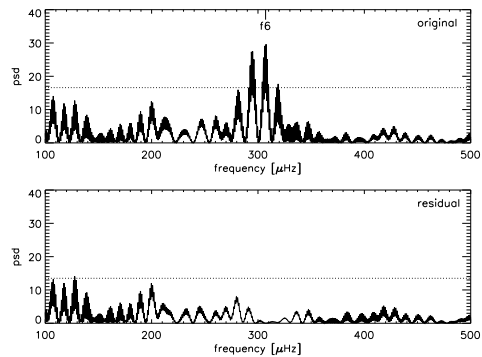}
\caption{LSPs of the $g$-mode regime of HS\,2201+2610 after subtraction 
of five $p$-modes (top). Subtracting the $g$-mode frequency f6 results in 
the residual LSP (bottom). Horizontal lines indicate
3\,$\sigma$ confidence levels.}
\end{figure}

\acknowledgements R. Lutz thanks the organizers for 
providing financial support. Based on observations at the CAHA at Calar Alto,
operated by MPIA and CSIC. DFG travel grants DR 281/20-1 and 
SCHU 2249/3-1.

\end{document}